\documentclass[english,12pt,a4paper]{article}
\usepackage{babel,amssymb,amsmath,graphicx,hyperref}

\begin{document}

\begin{center}
{\LARGE Gravitational lensing and wormhole shadows}\\~\\
{\large M.A. Bugaev$^1$,  I.D. Novikov$^{2,3,4}$, S.V. Repin$^{5,2}$, A.A. Shelkovnikova$^5$}
\end{center}
\begin{center}
\textit{$^1$Moscow Institute of Physics and Technology, \\
                     9, Institutskiy per., town Dolgoprudnyi, Moscow region, 141700, Russia}
\end{center}
\begin{center}
    \textit{$^2$Astro-Space Center of P.N. Lebedev Physical Institute, \\
                        84/32, Profsoyusnaya str., Moscow, 117997, Russia}
\end{center}
\begin{center}
    \textit{$^3$The Niels Bohr International Academy, The Niels Bohr Institute, \\
                        Blegdamsvej 17, DK-2100, Copenhagen, Denmark}
\end{center}
\begin{center}
    \textit{$^4$National Research Center Kurchatov Institute, \\
                        1, Akademika Kurchatova pl., Moscow, 123182, Russia}
\end{center}
\begin{center}
    \textit{$^5$Physics and Mathematics Colledge No. 2007, 9, Gorchakova str., Moscow, 117042, Russia}
\end{center}

\bigskip

\begin{abstract}
         The problem of bending and scattering of light rays passing outside from the entrance to a wormhole with 
zero gravitational mass is considered. The process of ray capture by a wormhole as well as the process of 
formation of a shadow when illuminated by a standard screen is investigated. These mechanisms are also 
compared to the case of motion of light rays in the vicinity of the Schwarzschild black hole.
\end{abstract}


     PASC number:  98.70.Ve

\section{Introduction}
         Recently, issues related to the physics of wormholes (WH) have attracted a special attention of specialists. 
This attention has increased after it was hypothesized that some galactic nuclei may not be the supermassive black 
holes (BH), but entrances to wormholes~\cite{Kardashev_2006, Nandi_2006}. There is a need to test this 
hypothesis.~\cite{Novikov_2021a}. A way to do so is to study the behavior of light rays in the vicinity of 
the entrance of a wormhole, as well as to analyse the possibility of the appearance of a shadow, similar to the case 
of BH, where shadow formation does take place. All of these questions are also of fundamental importance for 
the theory of wormholes. In this article we begin to explore these fundamental issues. We'll start with the simplest 
models. 

       In this paper, we will consider the simplest zero-mass Ellis-Bronnikov-Morris-Thorne wormhole model 
\cite{Ellis_1973, Bronnikov_1973, Morris_1988a, Morris_1988b}. Let us study the distortion of the motion of light 
rays passing in the vicinity of the entrance to a wormhole. We will study the light ray propagation distortions
caused by the wormhole in the viciniyty of its entrance.  Followed by comparison of those distortions to the ones
caused by the gravitational field of the Schwarzschild  BH.  Next, we will construct the shadow of a wormhole 
illuminated by the simplest model of a light screen and compare the result once again to the case of the Schwarzschild 
black hole illuminated by the same screen.

     The results will reveal the most important features of the process. In what follows, we turn to more complex 
and realistic situations.

      We will compare the processes in WHs and BHs. In view of this, we will not consider the rays passing into 
the wormhole. We will assume that the WH is filled with an opaque substance and the rays crossing the throat 
of the WH are absorbed by this substance.

       The implementation of our project consists of calculating the trajectories of light beams, i.e. zero geodesics 
in wormhole space-time. The corresponding geodesics have been calculated and discussed many times since 
the pioneering work \cite{Ellis_1973}, see also~\mbox{\cite{Novikov_2007}-\cite{Shaikh_2018}}. However, 
the problem requires a large amount of numerical simulation, and for this we needed other forms of the equations 
of motion and an analysis of their other properties.The results of this work are given in the Appendix. Of course, 
the properties of geodesics in Schwarzschild space-time are well known~\cite{Zeldovich_1967, Landau_1971}.

      In this paper, we do not consider the issues of WH instability (see \cite{Bronnikov_2013} about this) 
and assume that the WH metric does not depend on time.

\section{Bending of light rays}

      Consider in the WH metric:
\begin{equation}
    ds^2 = dt^2 - \cfrac{r^2}{r^2 - q^2}\,\, dr^2 - r^2
           \left(
              d\vartheta^2 + \sin^2\vartheta\, d\varphi^2
           \right)
\end{equation}
 the classical problem of wormhole ray scattering where light arrives from infinity with impact parameter~$b$.

\begin{figure}[htb]
  \centerline{
  \includegraphics[width=12cm]{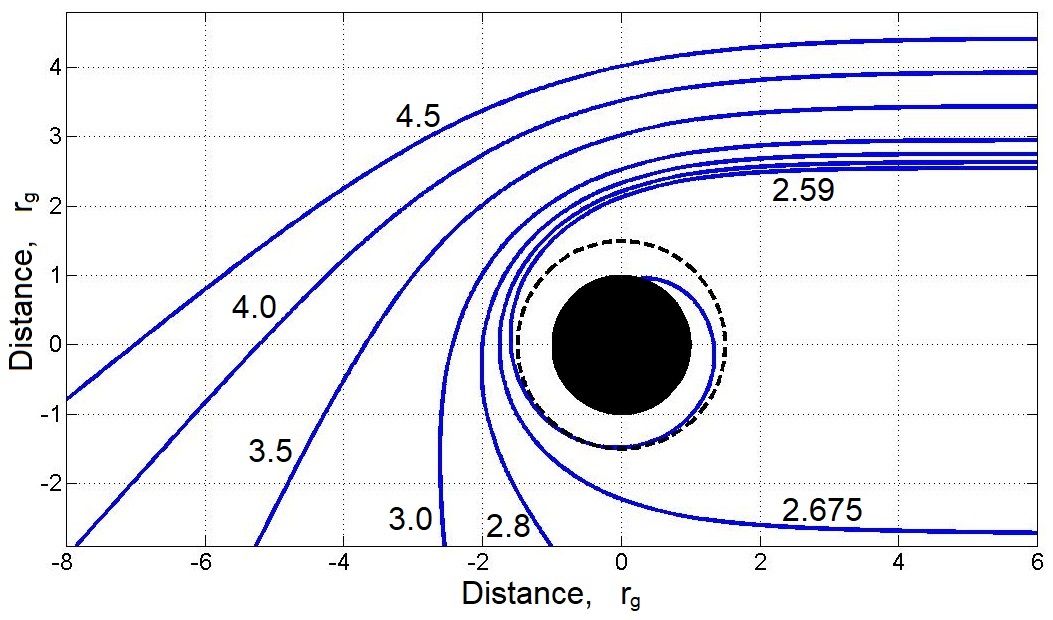}
             }
  \caption{Bending of the light rays by the Schwarzschild BH. A black circle with the radius of $r = r_g = 2GM/c^2$ 
                 is a black hole.  The dotted circle is $r = 3r_g/2$.
                 $b_{crit} = 3\sqrt 3\, r_g/2$  is the critical impact parameter of the BH beam capture.
                 The impact parameters are indicated for each trajectory.
                 The trajectory with an impact parameter $b = 2.59\, r_g$ enters the event horizon vertically.}
  \label{Schw_trajectories}
\end{figure}

       Let us recall what this problem looks like in the Schwarzschild BH metric, see Fig.~\ref{Schw_trajectories}.       
       
\begin{figure}[htb]
  \centerline{
  \includegraphics[width=8cm]{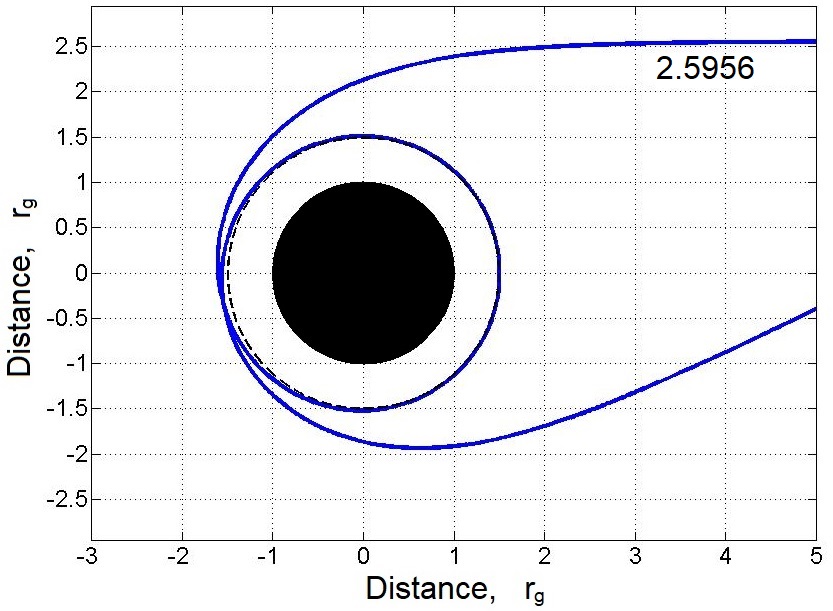}
             }
  \caption{Bending of the beam by the Schwarzschild black hole.  
               The value of the impact parameter is shown in the figure.
                The dotted circle has the radius $r =\cfrac 32\,\, r_g$.}
  \label{Schw_trajectories_2}
\end{figure}

        First of all, let us note that the BH is encompassed  by a sphere influence of outer radius $r = \cfrac{3}{2}\,\,r_g$,
around which the photons can move. Beams with the impact parameter $b_{crit}=\cfrac{3\sqrt 3}{2}\,\, r_g$ 
are captured by the black hole.  The beams with a slightly larger impact parameter, undergo bending and can go 
around  the BH many times in a tight spiral close to the circle $r=3r_g / 2$ before going back to infinity, 
see Fig.~\ref{Schw_trajectories_2}.
       
        Let us now turn to the curvature of light rays near the Ellis-Bronnikov-Morris-Thorne wormhole. At first 
glance, the overall picture should be fundamentally different from the classical one above, because in this case 
the~WH is massless, i.e. $m = 0$, there are no gravitational forces.Therefore, the classical utterance `` \dots 
under the influence of the gravitational field, the light beam is bent '' \cite{Landau_1971} would be inappropriate here. 
In this case, the light beam is bent due to the three-dimensional bending of space itself. This curvature of space, that 
solely determines ray bending, is analyzed in detail in~\cite{Novikov_2021b}. 
       
\begin{figure}[tbhp]
  \centerline{
  \includegraphics[width=8cm]{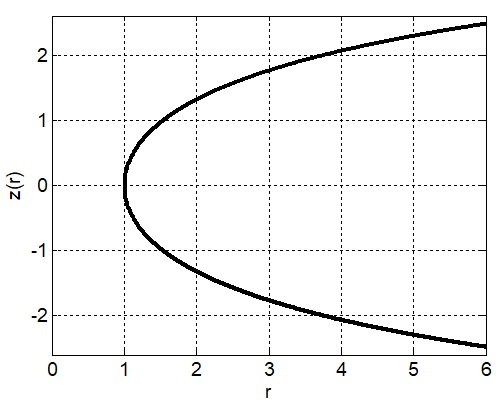}
             }
  \caption{The curve whose rotation about the $y$-axis forms the surface of the equatorial section of a WH. 
                The distances are normalized to the radius of the WH throat.}
  \label{Surface_MT}
\end{figure}

      Fig.~\ref{Surface_MT}, taken from the paper \cite{Novikov_2021b}, shows a curve whose rotation around 
the $z$-axis gives the surface of the equatorial section of the WH.

      The derivation of the equations of zero geodesics for WH and BH Schwarzschild metrics is given in the Appendix.

\begin{figure}[tbhp]
  \centerline{
  \includegraphics[width=13cm]{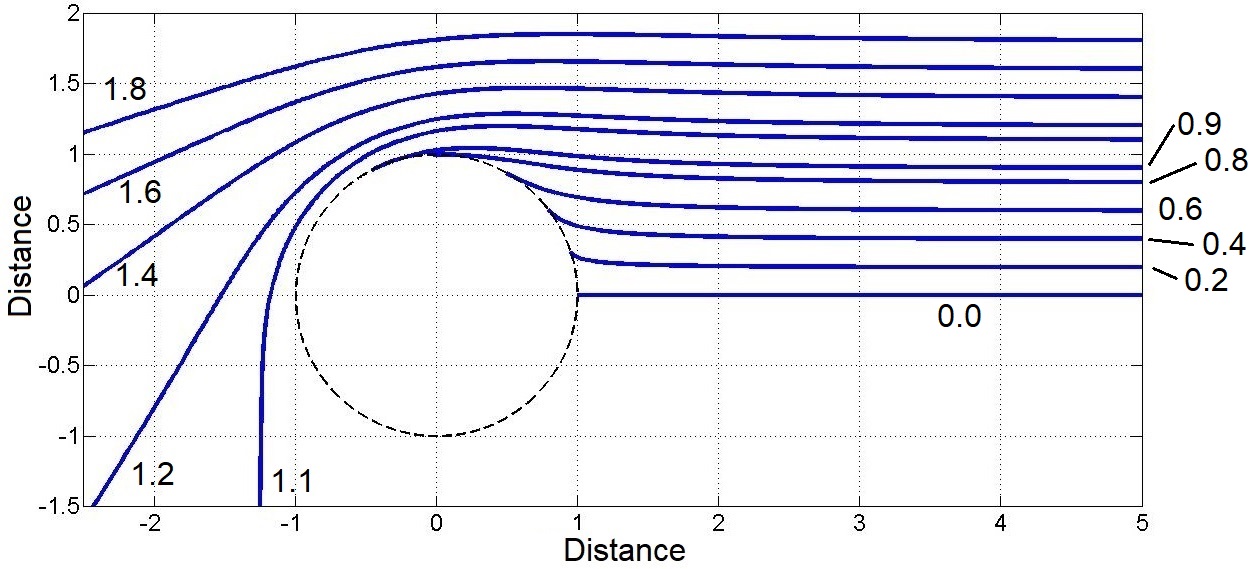}
             }
  \caption{Trajectories of light beans near a WH. Distances are measured in the units of WH throat radius. 
                 The throat of a wormhole is shown with a dashed line.
                 The impact parameter value is indicated for each trajectory.}
  \label{MT_trajectories_1}
\end{figure}

\begin{figure}[tbh]
  \centerline{
  \includegraphics[width=12cm]{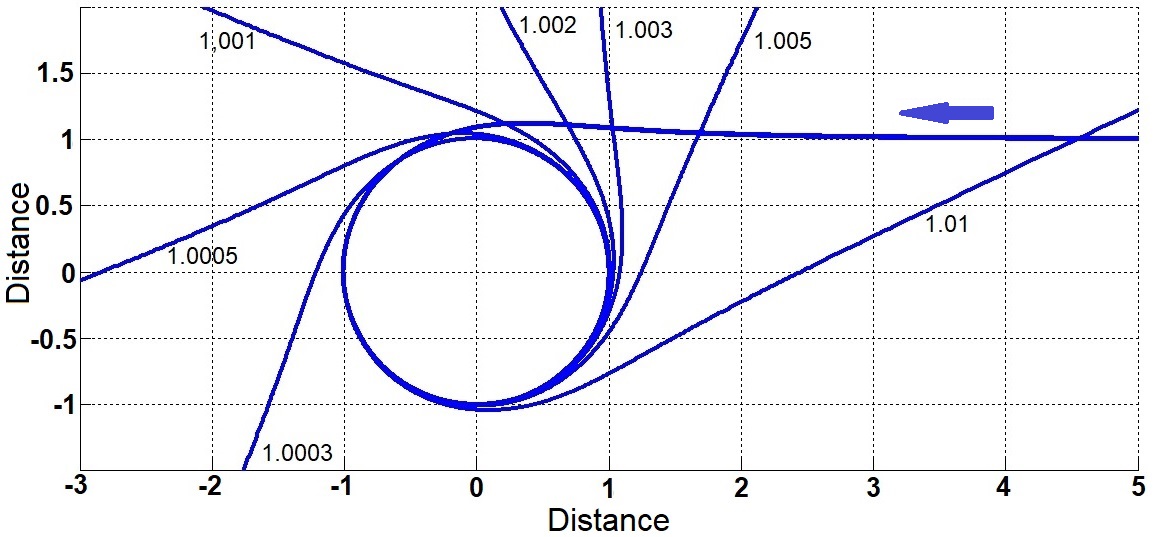}
             }
  \caption{Trajectories of light beams with a small impact parameter near the WH. 
                 Distances are measured in the units of WH throat radius.
                The impact parameter is indicated for each trajectory.}
  \label{MT_trajectories_2}
\end{figure}

        The trajectories in the vicinity of a WH are shown in Fig.~\ref{MT_trajectories_1}.  In this case, the circular 
orbits of light rays are also present. They are located at the throat of the WH at $r = q$.  As shown by 
computational analysis, the rays that arrive with an impact parameter less than $b_{crit} = q$ are captured by 
the wormhole and, according to our assumption, undergo absorbtion within it. Similar to the case of a BH, 
the trajectories that revolve multiple times around a WH do exist.  Examples are shown in 
Fig.~\ref{MT_trajectories_2}. These trajectories are important when constructing the~WH shadow.

      As it follows from the above results, despite the absence of gravitational forces, the picture of curved light 
rays is qualitatively similar to the case of a BH, although it differs numerically.

\section{Wormhole shadows}

       The shadows created by black holes against the background of various luminous formations are being studied 
in detail in modern theoretical astrophysics. Moreover, these shadows were recently discovered by astrophysical 
observations (\cite{EHT_collaboration_2019a}--\cite{EHT_collaboration_2019f}). All these works are of great 
importance both for the theory and for the experimental studies of the Universe.
\begin{figure}[tbh]
  \centerline{
  \includegraphics[width=12cm]{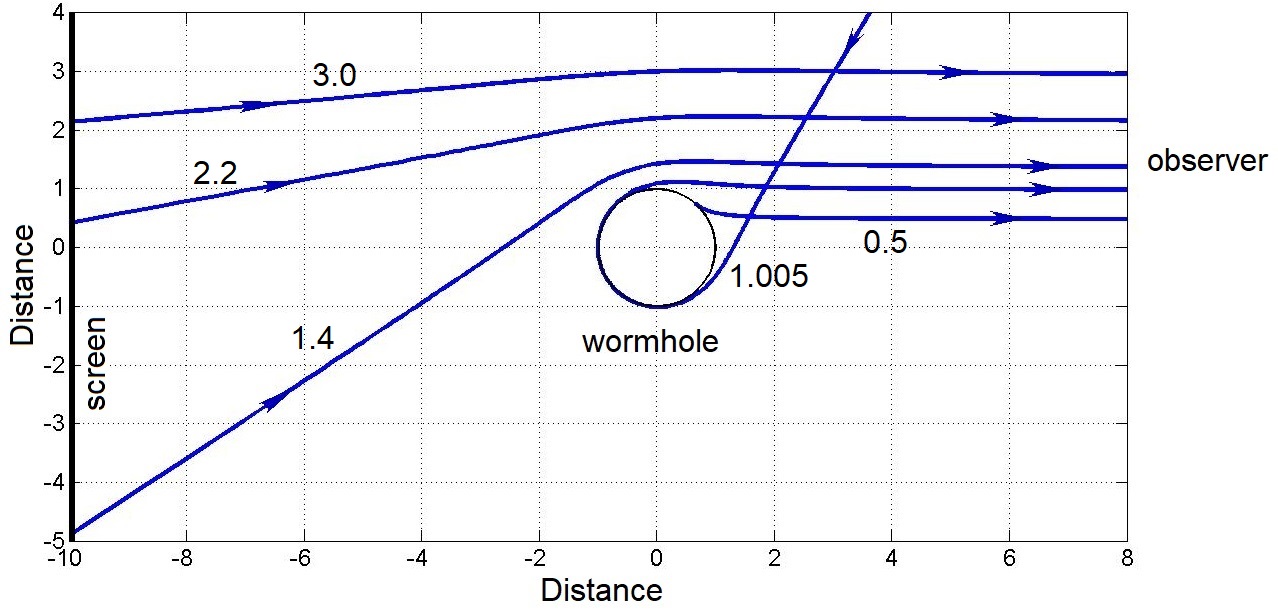}
             }
  \caption{Scheme for omputing the formation of the WH shadow when illuminated by a remote screen (shown in 
                the figure on the left, coordinate $-10$). The observer is on the far right.
                The distances are measured in units of the radius of the WH throat.  For each trajectory, the value of 
                the impact parameter is indicated.}
  \label{MT_trajectories_3}
\end{figure}

\begin{figure}[tbh]
  \centerline{
  \includegraphics[width=8cm]{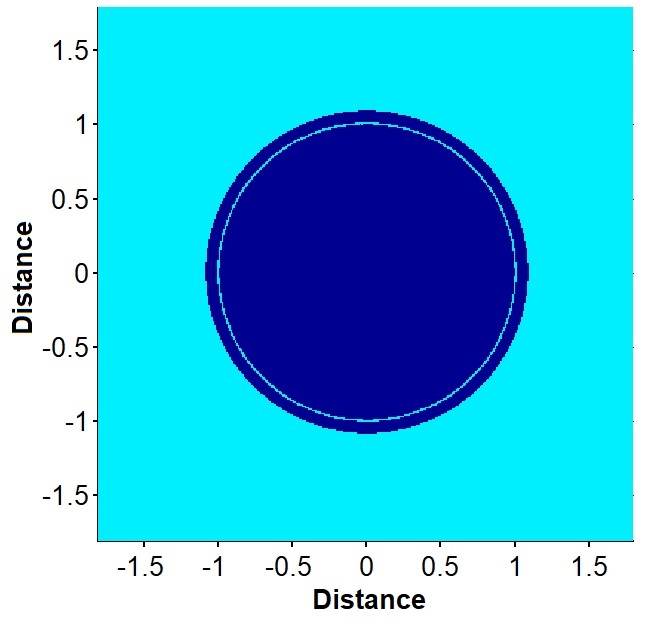}
  \includegraphics[width=8cm]{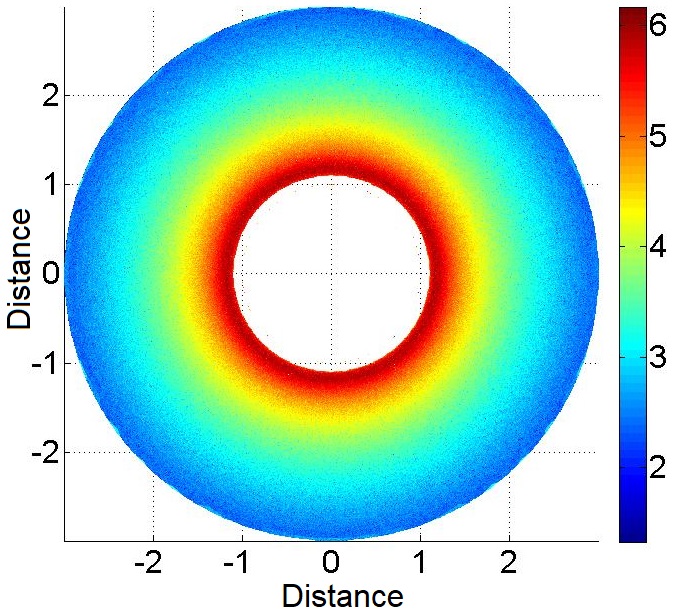}
             }
  \caption{The shape of the WH shadow and the distribution of the radiation intensity near its edge.
                 The radial coordinate is expressed in units of the radius of the WH throat.}
  \label{WH_MT_shadow}
\end{figure}

        The construction of the theory of WH shadows is of great importance for the detection of WHs in the Universe.
In this paper, we will consider the simplest examples of the theoretical computation of the structure of the 
WH shadow. We will consider the WH against the background of a distant infinite screen, radiating uniformly in 
all directions. Such a screen is called the Lambert source.

      Recall that we consider particular rays that move outside the WH, assuming that the interior of the WH is 
impenetrable for light. The results of Section 2 can be used to determine the structure the shadow that is formed.

       Fig.~\ref{MT_trajectories_3} shows a diagram of the corresponding computation. The parallel bundle of zero 
geodesic trajectories on the right is the trajectories that arrive to a distant observer. The observer sees only 
those trajectories that start on the screen. Other trajectories either come from infinity or from the wormhole itself, 
and then carry no light.  Considering these latter trajectories, the shadow of the wormhole can be constructed.
The image in Fig.~\ref{MT_trajectories_3} is a reversal in time of the pictures in 
Fig.~\ref{MT_trajectories_1}--\ref{MT_trajectories_2}.

\begin{figure}[htb]
  \centerline{
  \includegraphics[width=8cm]{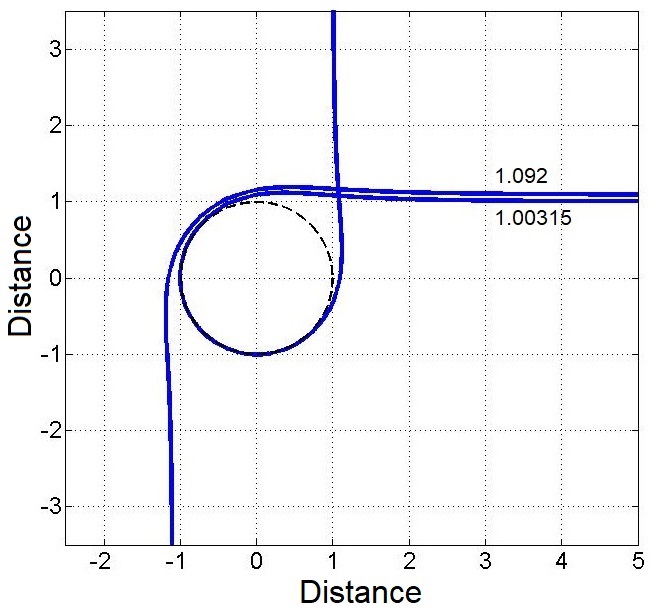}
             }
  \caption{Critical trajectories in the WH field.
                 The radial coordinate is expressed in units of the radius of a wormhole throat.}
  \label{WH_MT_trajectories_crit}
\end{figure}

\begin{figure}[tbh]
  \centerline{
  \includegraphics[width=8cm]{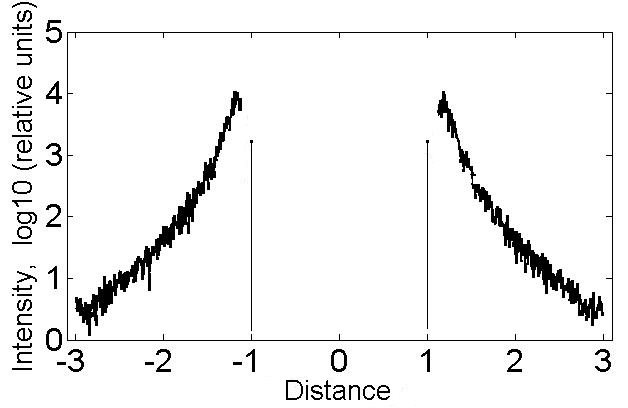}
             }
  \caption{Radiation intensity distribution in the equatorial section of the WH shadow.}
  \label{MT_intens_plot}
\end{figure}

        Fig.~\ref{WH_MT_shadow} shows the shadow of the WH and the distribution of the radiation intensity 
near its boundary: so called corona or aura. The shadow is a circle with rings (an infinite number of rings) near 
the inner edge of the shadow. Fig.~\ref{WH_MT_trajectories_crit} shows the trajectories of zero geodesics defining 
the edge of the WH shadow and the first ring, i.e. the trajectories turning $90^\circ$ and $270^\circ$, respectively.
The impact parameters of these trajectories are as follows:
\begin{equation}
        b_{90} = 1.092q, \qquad \qquad  b_{270} = 1.00315q,
\end{equation}
where $q$ as before, denotes the radius of the WH throat. The subsequent rings correspond to the smaller values of 
the impact parameter, up to $b_{crit} = q$.

\begin{figure}[tbh]
  \centerline{
  \includegraphics[width=8cm]{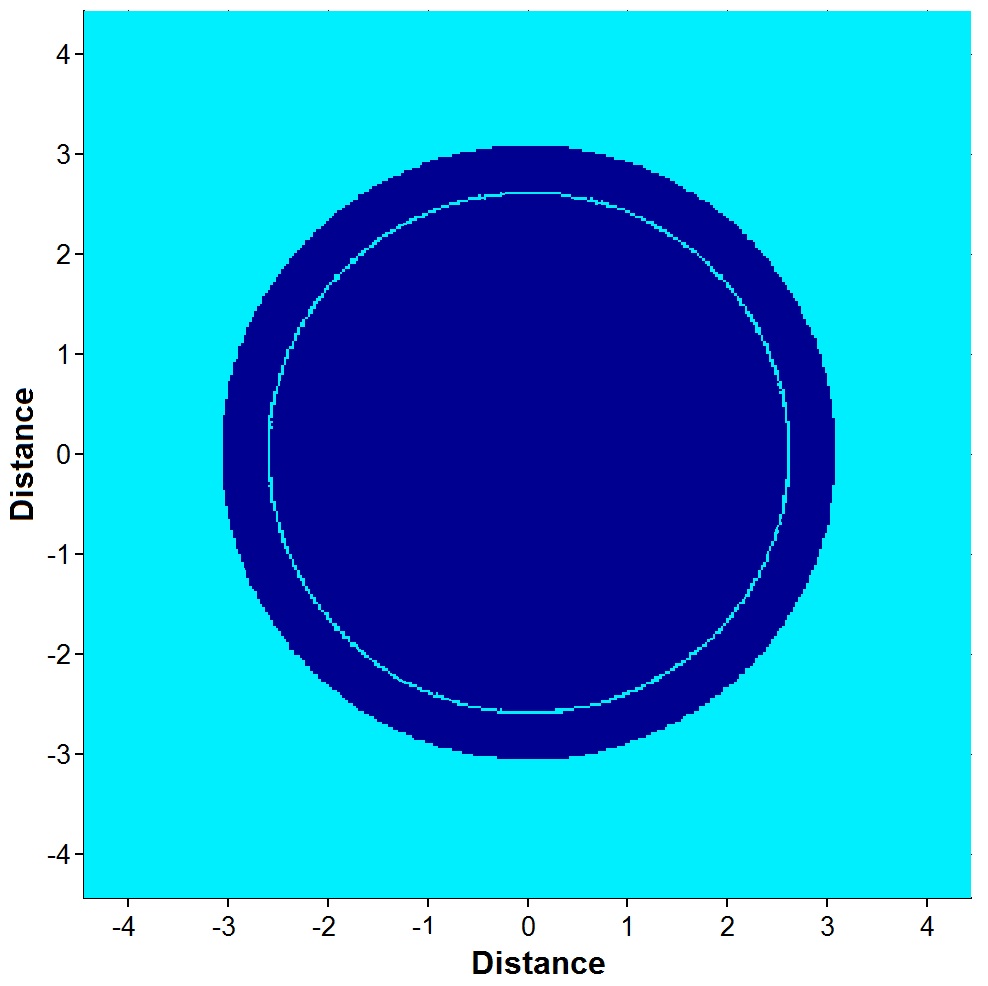}
  \includegraphics[width=8cm]{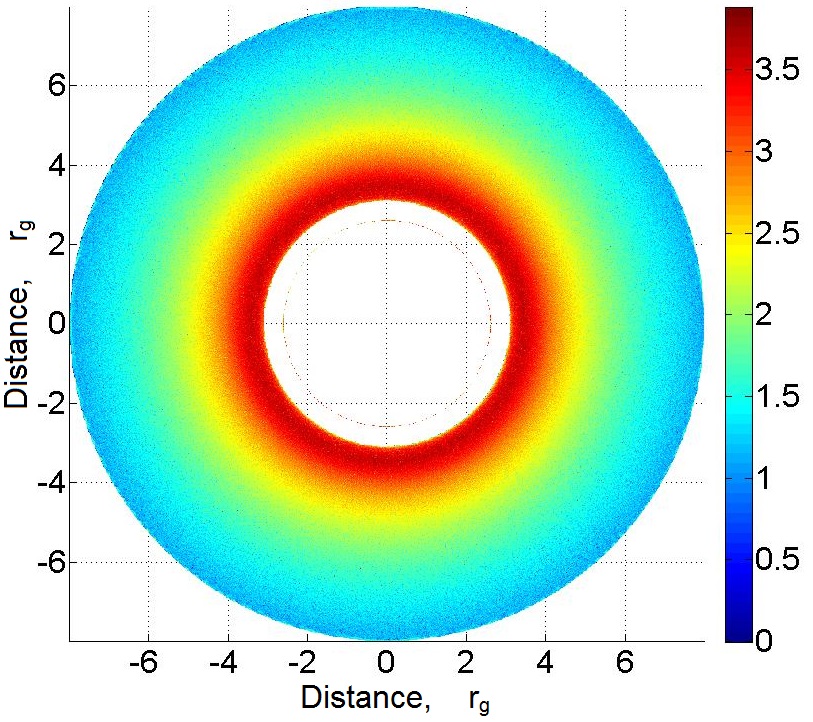}
             }
  \caption{The shape of the shadow of the Schwarzschild black hole and the distribution of the radiation 
                intensity near its edge. The radial coordinate is expressed in units of the gravitational radius 
                $r_g = 2Gm / c^2$.}
  \label{Schw_shadow}
\end{figure}

        The distribution of the radiation intensity in the equatorial section of the WH shadow is shown in
 Fig.~\ref{MT_intens_plot}. When approaching the shadow boundary, the radiation intensity in the corona increases 
 by several orders of magnitude and turns out to be nonlinear on a logarithmic scale.Unfortunately, we cannot compare 
 the intensity of the inner ring radiation with the intensity of the corona due to the limited accuracy of the computations.
 This requires other numerical methods.
        
        For comparison with Fig.~\ref{WH_MT_shadow}, Fig.~\ref{Schw_shadow} shows the Schwarzschild black 
hole shadow and the distribution of the radiation intensity in its corona against the background of the same bright 
screen. As follows from Fig.~\ref{Schw_shadow}, the position of the bright ring inside the shadow silhouette is very 
different for the BH and the WH, and this may be one of the signs by which these objects can be distinguished from 
each other.

\begin{figure}[tbh]
  \centerline{
  \includegraphics[width=8cm]{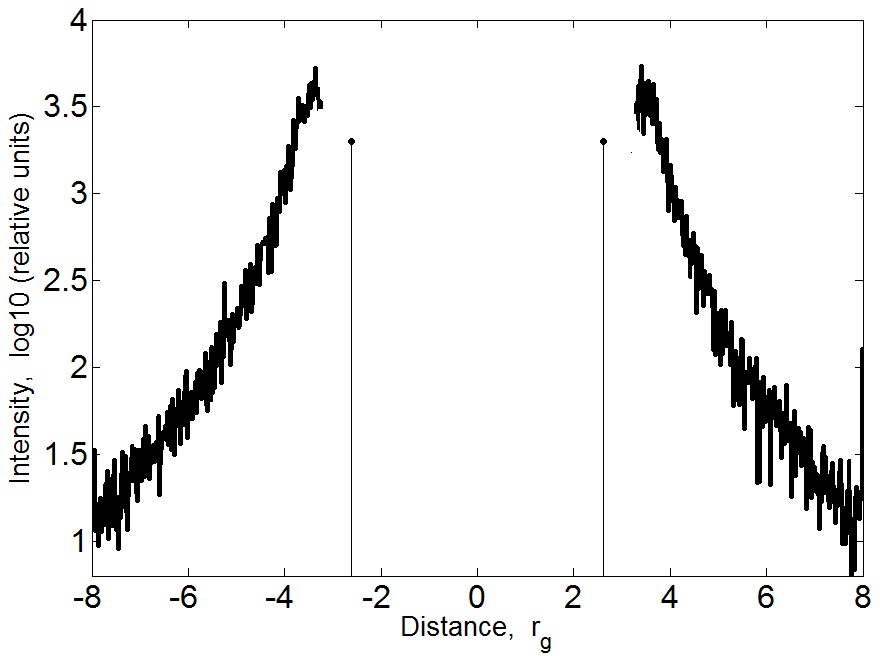}
             }
  \caption{The distribution of the radiation intensity in the equatorial section of the shadow of the Schwarzschild 
                black hole.}
  \label{Schw_intens_plot}
\end{figure}

        Another sign may be the intensity distribution in the equatorial section of the shadow of the WH and BH.
For the latter, the distribution is shown in Fig.~\ref{Schw_intens_plot} and can be compared with 
Fig.~\ref{MT_intens_plot}. The increase in the intensity of the corona as it approaches the boundary of the shadow 
turns out to be more gentle for the Schwarzschild BH.  This fact can also be used in observations in order to distinguish 
a BH from a WH.

\section{Conclusion}

       We emphasize that although the physical properties of the WH and BH metrics differ sharply from each other 
(there is no gravitation in the WH metric at all), the lensing of light rays and the shape of the shadow they create 
is similar to each other, differing mainly in numerical parameters.

     As noted in Section I, in our further work we will consider more complex models of the WH and screens.

\section{Acknowledgments}

     The authors are grateful to I.D. Novikov Jr. for technical support. 
S.R. is grateful to R.E.~Beresneva, O. N. Sumenkova and O.A. Kosareva for the opportunity to work fruitfully 
on this problem. 

\section{Appendix. Equations of motion of a photon}

    The metric of the WH can be written in the form:
\begin{equation}
    ds^2 = dt^2 - dR^2 - \left(R^2 + q^2\right) 
           \left(
              d\vartheta^2 + \sin^2\vartheta\, d\varphi^2
           \right),
\end{equation}
or
\begin{equation}
    ds^2 = dt^2 - \cfrac{r^2}{r^2 - q^2}\,\, dr^2 - r^2
           \left(
              d\vartheta^2 + \sin^2\vartheta\, d\varphi^2
           \right)
\end{equation}
where
\begin{equation}
    r^2(R) = R^2 + q^2, 
\end{equation}
$q$ is the size of WH throat, and the radial coordinate $r$ is chosen so that the circumference is $2 \pi r$.           

        Hamilton--Jacobi equation 
\begin{equation}
     g^{ik}
     \cfrac{\partial S}{\partial x^i}
     \cfrac{\partial S}{\partial x^k} - m^2 = 0
\end{equation}
for the wormhole metric can be written as:
\begin{equation}
     \left(\cfrac{\partial S}{\partial t}\right)^2 -
     \cfrac{r^2 - q^2}{r^2} \left(\cfrac{\partial S}{\partial r}\right)^2 -
     \cfrac{1}{r^2} \left(\cfrac{\partial S}{\partial \vartheta}\right)^2 -
     \cfrac{1}{r^2\sin^2\vartheta}\left(\cfrac{\partial S}{\partial \varphi}\right)^2 -
     m^2 = 0.
     \label{HJ_equation}
\end{equation}

        By analogy with the Schwarzschild metric, the solution is sought in the form:
\begin{equation}
         S = -Et + L\varphi + S_r(r) + S_\theta(\theta),
\end{equation}
since the coordinates $t$ and $\phi$ are cyclic, i.e. are not explicitly included in the metric tensor and 
the Hamilton - Jacobi equation. In the equation (\ref{HJ_equation}), the variables can be separated and we get 
the equations of motion:
\begin{eqnarray}
      \cfrac{dt}{d\lambda} & = & Er^2, \label{Eq_motion_1}  \\
      \left(\cfrac{dr}{d\lambda}\right)^2 & = & \left(E^2 - m^2\right) r^4 - 
                                  \left(q^2 \left(E^2 - m^2\right) + Q + L^2 \right) +
                                  q^2 \left(Q + L^2 \right), \label{Eq_motion_2} \\
     \left( \cfrac{d\theta}{d\lambda}\right)^2 & = & Q - \cfrac{L^2\cos^2\theta}{\sin^2\theta}\,\,, 
                                 \label{Eq_motion_3} \\
      \cfrac{d\varphi}{d\lambda} & = & \cfrac{L}{\sin^2\theta}\,\,,  \label{Eq_motion_4}
\end{eqnarray}
where the Carter separation constant $K$ according to \cite{Misner_1973} is written in the form: $K = Q + L^2$,
which differs from its recording in \cite{Landau_1971}. The equations of motion for a quantum are obtained by 
putting $m = 0$ in the system (\ref{Eq_motion_1}) -(\ref{Eq_motion_4}). However, in this case the equations of 
motion do not depend on three constants: $E$, $L$ and $Q$, but on two Chandrasekhar constants: 
$\xi = Q / q^2E^2$ and $\eta = L / qE$. In addition, the equations (\ref{Eq_motion_2})-(\ref{Eq_motion_3}) 
contain a square root function, which is inconvenient for serial computations. For the convenience of computlations, 
each of these equations can be replaced by two equations, thereby increasing the order of the system 
\cite{Zakharov_1994, Zakharov_1999}. Finally, the system of equations for the motion of a quantum in 
the WH metric can be written as:
\begin{eqnarray}
      \cfrac{dt}{d\sigma} & = & \cfrac{1}{r^2}\,\,, \label{Eq_motion2_1}  \\
      \cfrac{dr}{d\sigma} & = & r_1\,, \label{Eq_motion2_2}  \\
      \cfrac{dr_1}{d\sigma} & = & 2 \left(\eta - \xi^2\right) r^3 - \left(1 + \eta + \xi^2\right) r\,, 
                                \label{Eq_motion2_3} \\
     \cfrac{d\theta}{d\lambda} & = & \theta_1\,, \label{Eq_motion2_4} \\ 
     \cfrac{d\theta_1}{d\lambda} & = & \cfrac{\xi^2\cos\theta}{\sin^3\theta}\,\,,\label{Eq_motion2_5} \\
      \cfrac{d\varphi}{d\lambda} & = & \cfrac{\xi}{\sin^2\theta}\,\,.  \label{Eq_motion2_6}
\end{eqnarray}

\begin{figure}[tbh]
  \centerline{
  \includegraphics[width=6cm]{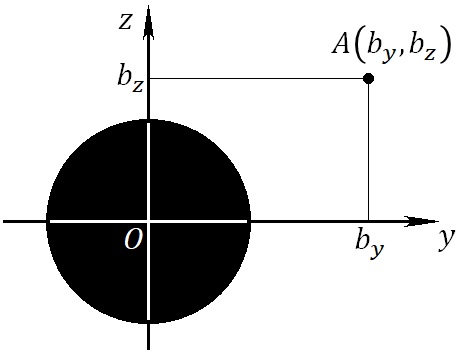}
             }
  \caption{Impact parameter for a quantum directed towards a wormhole.}
  \label{WH_impact_param}
\end{figure}

       To set the initial values, assume that the observer is at a point with coordinates 
$(r,\theta,\varphi) = (r_0, \theta_0, 0)$ (i.e., in the plane $xz$ in Cartesian coordinates) and launches a quantum 
at the direction of the WH to the point with coordinates $A\left(b_y, b_z \right)$ in the plane of the sky and 
the impact parameter $b = \sqrt{b_y^2 + b_z^2}$. The geometrical location of this point is illustrated in 
Fig.~\ref{WH_impact_param}. Then the Chandrasekhar constants can be expressed explicitly through 
the coordinates of the observer and the impact parameter:
\begin{equation}
       \xi = \cfrac{r_0 b_y \sin\theta_0}{\sqrt{r_0^2 + b_y^2 + b_z^2}}\,\,,  \qquad
       \eta = \cfrac{r_0^2 \left(b_z^2 +b_y^2\cos^2\theta_0\right)}{r_0^2 + b_y^2 + b_z^2}\,\,.
            \label{xi_eta_MT}
\end{equation}
All linear dimensions in the formulas (\ref{xi_eta_MT}) are normalized to the radius of the WH throat, 
i.e.in Fig.~\ref{WH_impact_param} $b_y\approx 2$, $b_z\approx 1.5$.

        For the Schwarzschild metric, a similar system of equations is written as:
\begin{eqnarray}
      \cfrac{dt}{d\sigma} & = & \cfrac{1}{r^2 \left(1 - 2r\right)}\,\,, \label{Eq_motion3_1}  \\
      \cfrac{dr}{d\sigma} & = & r_1\,, \label{Eq_motion3_2}  \\
      \cfrac{dr_1}{d\sigma} & = & 3 \left(\eta + \xi^2\right) r^2 - \left(\eta + \xi^2\right) r =
                           \left(\eta + \xi^2\right)\left(3r - 1\right) r\,, 
                                \label{Eq_motion3_3} \\
     \cfrac{d\theta}{d\lambda} & = & \theta_1\,, \label{Eq_motion3_4} \\ 
     \cfrac{d\theta_1}{d\lambda} & = & \cfrac{\xi^2\cos\theta}{\sin^3\theta}\,\,,\label{Eq_motion3_5} \\
      \cfrac{d\varphi}{d\lambda} & = & \cfrac{\xi}{\sin^2\theta}  \label{Eq_motion3_6}
\end{eqnarray}
and the Chandrasekhar constants
\begin{equation}
       \xi = \cfrac{r_0^{3/2} b_y \sin\theta_0}
                         {\sqrt{\left(r_0 - 2\right)\left(r_0^2 + b_y^2 + b_z^2\right)}}\,\,,  \qquad
       \eta = \cfrac{r_0^3 \left(b_z^2 +b_y^2\cos^2\theta_0\right)}
                            {\left(r_0 - 2\right)\left(r_0^2 + b_y^2 + b_z^2\right)}\,\,.
            \label{xi_eta_Schw}
\end{equation}
        
       For the numerical solution of the ordinary differential equations, there are many software packages that are freely 
distributed on the Internet. The calculations were performed with a relative accuracy $\delta = 10^{- 7}$ and were 
verified using the first integrals of the systems (\ref{Eq_motion2_1})--(\ref{Eq_motion2_6}) and (\ref{Eq_motion3_1})--(\ref{Eq_motion3_6}) \cite{Zakharov_1994, Zakharov_1999}.


\bigskip

\begin{thebibliography}{99}
\bibitem{Kardashev_2006} N.S. Kardashev, I.D. Novikov, A.A. Shatskiy, Astronomy Reports, 
  \textbf{50}, 601 (2006)
\bibitem{Nandi_2006}  K.K. Nandi, Y-Zh. Zhang, A.V. Zakharov, Phys. Rev. D, \textbf{74}, 024020 (2006)
\bibitem{Novikov_2021a} I.D. Novikov, S.F. Likhachev, Yu.A. Shchekinov, A.S. Andrianov et al.,
   Physics--Uspekhi, \textbf{64}  (2021)
\bibitem{Ellis_1973} Ellis H, \textit{J Math Phys} \textbf{14}, 104 (1973)
\bibitem{Bronnikov_1973} Bronnikov K A, Acta Phys Pol \textbf{84} 251 (1973)
\bibitem{Morris_1988a} Morris M S, Thorne K S \textit{Am. J. Phys.} \textbf{56}, 395 (1988)
\bibitem{Morris_1988b}  M.S. Morris, K.S. Thorne, U. Yurtsever,  \textit{Phys. Rev. Lett.} \textbf{61}, 1446 (1988)
\bibitem{Novikov_2007} I.D. Novikov, N.S. Kardashev, A.A. Shatskii, Physics--Uspekhi, \textbf{50}, 965 (2007)
\bibitem{Shatskiy_2008} A.A. Shatskii, I.D. Novikov, N.S. Kardashev,  Physics--Uspekhi,  \textbf{51}, 457 (2008)
\bibitem{Shatskiy_2009} A.A. Shatskii,  Physics--Uspekhi,  \textbf{52}, 811 (2009)
\bibitem{Nedkova_2013} P.G. Nedkova, V. Tinchev,  S.S. Yazadjiev, Phys. Rev. D \textbf{88}, 124019 (2013)
\bibitem{Shatskiy_2015} A.A. Shatskiy, Yu.Yu. Kovalev, I.D. Novikov,  
              Journal of Experimental and Theoretical Physics, \textbf{120}, 798 (2015)
\bibitem{Ohgami_2015} T. Ohgami, N. Sakai, Phys. Rev. D, \textbf{91}, 124020 (2015)
\bibitem{Abdujabbarov_2016} A. Abdujabbarov, B. Juraev, B. Ahmedov, Z. Stuchlik, 
             Astrophys. Space Sci. \textbf{361}, 226 (2016)
\bibitem{Shaikh_2018} R. Shaikh, Phys. Rev. D \textbf{98}, 024044 (2018)
\bibitem{Zeldovich_1967}  Ya.B. Zeldovich, I.D. Novikov. Relativistic Astrophysics, Moscow, Nauka, 1967, 
\bibitem{Landau_1971}  L. D. Landau, E. M. Lifshitz (1971). \textit{The Classical Theory of Fields}. 
   Vol. 2 (3rd ed.). Pergamon Press. ISBN 978-0-08-016019-1.)
\bibitem{Bronnikov_2013}  K.A. Bronnikov, L.N. Lipatova, I.D. Novikov, A.A. Shatskiy, Gravitation and Cosmology,
   \textbf{19}, 269 (2013)
\bibitem{Novikov_2009} D.I. Novikov, A.G. Doroshkevich, I.D. Novikov, A.A. Shatskii, 
   Astronomy Reports, \textbf{53}, 1079 (2009) 
\bibitem{Shinkai_2002} H.-A. Shinkai,  S.A. Hayward, Phys. Rev. D, \textbf{66}, 044005 (2002);  gr-qc/0205041
\bibitem{Doroshkevich_2009} A.G. Doroshkevich, J. Hansen, I.D. Novikov, A.A. Shatskiy, 
    Int. J. Mod. Phys. D, \textbf{18}, 1665 (2009)
\bibitem{Novikov_2021b} I.D. Novikov, S.V. Repin, Astronomy Reports, \textbf{65}, 1  (2021)
\bibitem{Misner_1973} Misner C.W., Thorne K.S., Wheeler J.A., \textit{Gravitation}, 
   San Francisco: W. H. Freeman, ISBN 978-0-7167-0344-0.
\bibitem{EHT_collaboration_2019a} Event Horizon Telescope Collaboration, 
   Astrophys. J. Lett., \textbf{875}, L1 (2019);  arXiv 1906.11238
\bibitem{EHT_collaboration_2019b} Event Horizon Telescope Collaboration, 
   Astrophys. J. Lett., \textbf{875}, L2 (2019);  arXiv 1906.11239
\bibitem{EHT_collaboration_2019c} Event Horizon Telescope Collaboration, 
   Astrophys. J. Lett., \textbf{875}, L3 (2019);  arXiv 1906.11240
\bibitem{EHT_collaboration_2019d} Event Horizon Telescope Collaboration, 
   Astrophys. J. Lett., \textbf{875}, L4 (2019);  arXiv 1906.11241
\bibitem{EHT_collaboration_2019e} Event Horizon Telescope Collaboration, 
   Astrophys. J. Lett., \textbf{875}, L5 (2019);  arXiv 1906.11242
\bibitem{EHT_collaboration_2019f} Event Horizon Telescope Collaboration, 
   Astrophys. J. Lett., \textbf{875}, L6 (2019);  arXiv 1906.11243
\bibitem{Zakharov_1994} A.F. Zakharov, Mon. Not. of Royal Astron. Soc., \textbf{269}, 283 (1994)
\bibitem{Zakharov_1999}  A.F. Zakharov, S.V. Repin, Astronomy Reports, \textbf{43}, 705 (1999)

\end{thebibliography}
\end{document}